\pdfoutput=1
\documentclass[cits]{JINST}
\let\ifpdf\relax

\DeclareFontFamily{U}{euc}{}
\DeclareFontShape{U}{euc}{m}{n}{<-6>eurm5<6-8>eurm7<8->eurm10}{}%
\DeclareSymbolFont{AMSc}{U}{euc}{m}{n} 
\DeclareMathSymbol{\umu}{\mathord}{AMSc}{"16}
\usepackage{caption}
\usepackage[pdftex]{graphicx}
\usepackage{amsmath}

\graphicspath{{figs/}}

\title{Results of the SDHCAL technological prototype.\\ \vspace{2 mm}{ \large Talk presented at the International Workshop on Future Linear Colliders (LCWS13), Tokyo, Japan, 11-15 November 2013.}}

\author{A.Steen\thanks{Universit\'{e} Claude Bernard Lyon 1, IPNL} , On behalf of the CALICE Collaboration.\\steen@ipnl.in2p3.fr}

\abstract{The SDHCAL technological prototype that has been completed in 2012 was exposed to beams of pions and electrons of different energies at the CERN SPS for a total time period of 5 weeks. The data has been analyzed within the CALICE collaboration. Preliminary results indicate that a highly granular hadronic calorimeter conceived for PFA application is also a powerful tool to separate pions from electrons. The SDHCAL provides also a very good resolution of hadronic showers energy measurement. The use of multi-threshold readout mode shows a clear improvement of the resolution at energies exceeding 30 GeV with respect to the binary readout mode. Simulations of the pion interactions in the SDHCAL are presented and new ideas to improve on the energy resolution using the topology of hadronic showers are mentioned.}

\begin{document}



\section{Introduction}
\label{sec:intro}
The CALICE Semi-Digital Hadronic CALorimeter (SDHCAL) prototype achieved in 2011 was built for several purposes. One of them is to show that highly granular gaseous calorimeters can provide a powerful tool  to measure hadrons energy and to apply the Particle Flow Algorithm (PFA). First studies using Hough transform method tends to confirm that the tracking capability of such detectors will be a useful tool for the PFA and for the estimation of the detector behaviour $\it{in\ situ}$ \cite{houghNote}. Another motivation for the SDHCAL was to build a detector compatible with the futur linear collider requirements in terms of efficiency, low power consumption and compactness. Imaging calorimeters could also bring interesting tools for the hadronic shower models study.
In the first section of this paper, a brief description of the SDHCAL prototype and the experimental set-up during test beams will be presented. Then, in section \ref{sec:building-perf}, the event building procedure and detector behaviour will be discussed. Hadronic energy resolution and the different methods to obtain it, will be discussed in the last section.

\subsection{The SDHCAL prototype}
\label{sec:intro-proto}
\begin{figure}[!ht]
  \begin{center}
    \includegraphics[width=.8\textwidth]{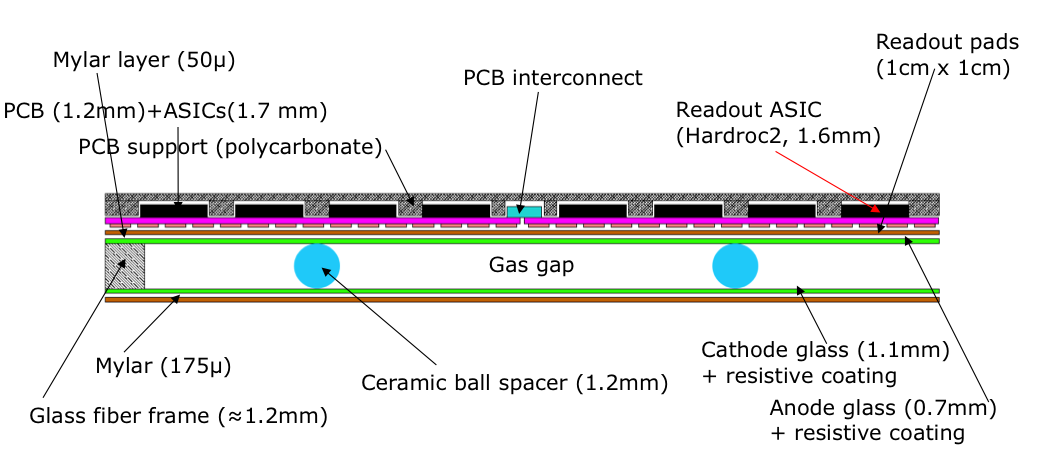}
    \caption{Schematic Glass Resistive Plate Chamber (GRPC) cross section.}
    \label{figs.cassette}
  \end{center}
\end{figure}

The SDHCAL prototype is a sampling calorimeter made of 48 active layers. Each layer is a 1 $m^2$ Resistive Plate Chambers (RPC). The cathode and the anode are glass plates with a thickness of 1.1 mm and 0.7 mm respectively. The gas gap between the two electrodes is 1.2 mm. 9216 pads of 1 $cm^2$ each compose the readout of one chamber. So, the number of electronic channels for the whole prototype is 442368. Readout signal from those pads is then digitzed and recorded in 2-bit format within HARDROC2 ASIC \cite{asic} located on the other face of the electronic board. Each chamber contains 144 ASICs read out by 3 Detector InterFace (DIF) cards. The aim of the thresholds is to have additionnal information about the number of charged particles which cross the pad. This will allow to reduce the saturation effect on the energy resolution at high energy. Chamber with its electronic board is put in a stainless steel cassette of 2.5 mm thickness per side. A cross section of one detector is shown in Fig.\ \ref{figs.cassette}.
The cassettes are then inserted between stainless steel absorber layers which compose a self supporting mechanical structure. Each absorber layer is 1.5 cm thick. So the absorber thickness between two following RPCs is 2 cm. This leads to a total depth of 6$\lambda_I$ for the SDHCAL prototype.

\subsection{Test beam and set-up}
\label{sec:intro-setup}
In order to validate the SDHCAL technology, the prototype was exposed to muons, pions and electrons at the CERN H2 beam line of the SPS in May and November 2012, and at the H6 beam line in August 2012. In the following, only data taken on the H6 SPS beam line are considered. To keep a high detection efficiency and take data in the SDHCAL optimal running condition, the different optics along the beam lines were set to enlarge as much as possible the beam size. Indeed, the GRPC detection efficiency is decreasing when the particle rate becomes too high \cite{rpcRate}. So, beam line optics were set to have a particle rate lower than 100 $Hz/cm^2$ (i.e. less than 1000 particles/spill). The detectors run in avalanche mode. The high voltage applied on the electrodes was 6.9 kV. The gas mixture was 93$\%$ of TetraFluoroEthane(TFE), 5$\%$ of $CO_2$ and 2$\%$ of $SF_6$. The three thresholds were set at 114 fC, 5 pC and 15 pC (0.1,4,12.5 MIP). During test beam, the trigger-less acquisition mode was used. In this mode, all hits are recorded until one ASIC memory is full. Then all ASICs stop their acquisition and allow the readout of the data by the DIFs. Acquisition restarts automatically at the end of the data transfer. The high amount of electronic channels leads to a heat production and associated noise. To reduce this noise, the power plused mode was used. With this mode, the detector electronics (DIF, ASIC) are switched off when there is no particle delivered by the beam line. At SPS, the spill duration is about 9 s within a cycle of 45 s. During those tests beam, no gain correction was applied. The same electronic gain was used for all the channels. Consequently, the following results show the performance of such a technology without any correction.


\section{Event building and detector behaviour}
\label{sec:building-perf}

\subsection{Event building}
\label{sec:building-perf-building}
\begin{figure}[!ht]
  \begin{center}
    \includegraphics[width=.5\textwidth]{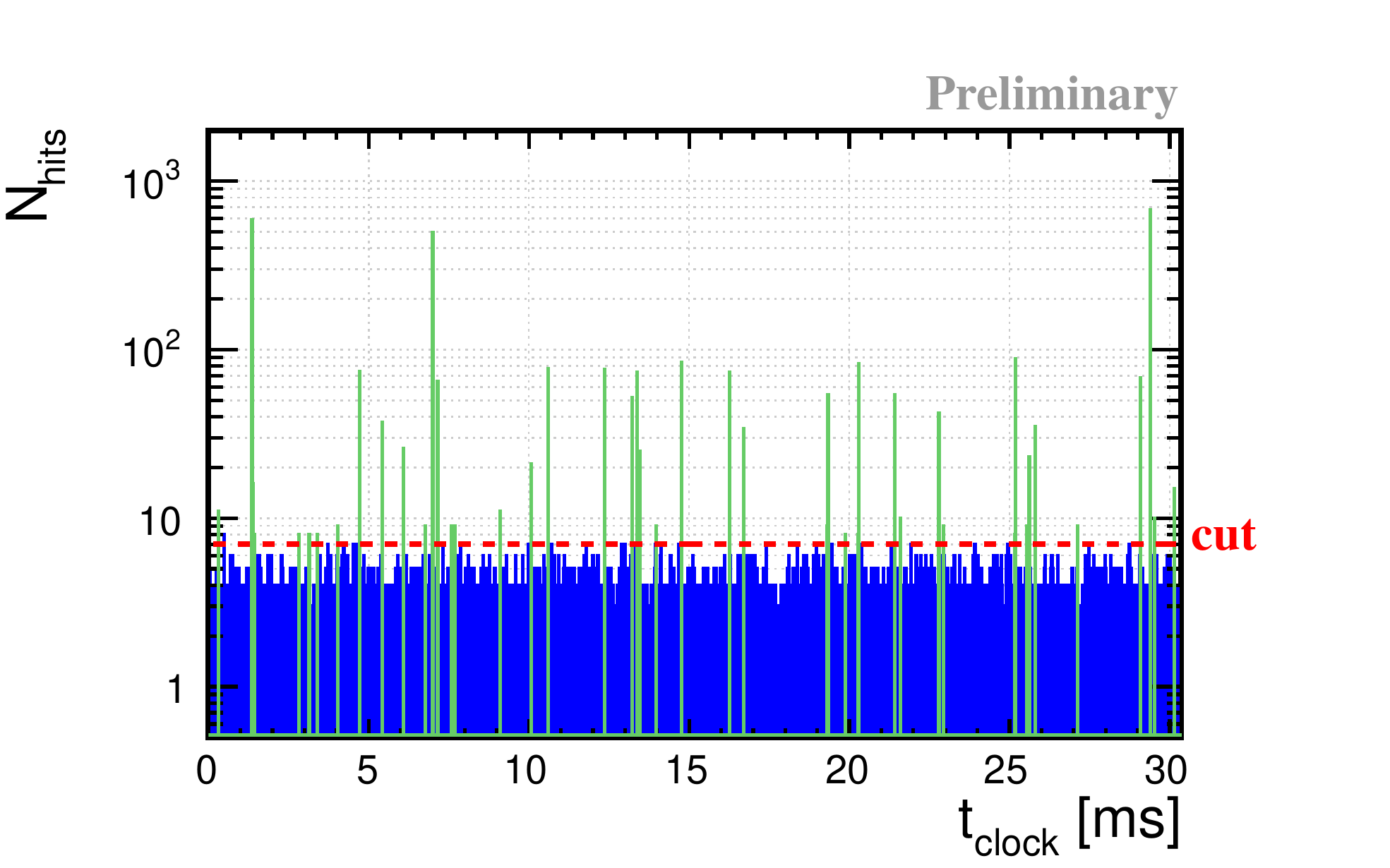}
    \caption{Hit time spectrum. Each bin corresponds to 200 ns. Bins in green ($\#_{hit}>7$) are the candidates to build a physical events.}
    \label{figs.time_spectrum}
  \end{center}
\end{figure}

\begin{figure}[!ht]
  \begin{center}
    \includegraphics[width=.5\textwidth]{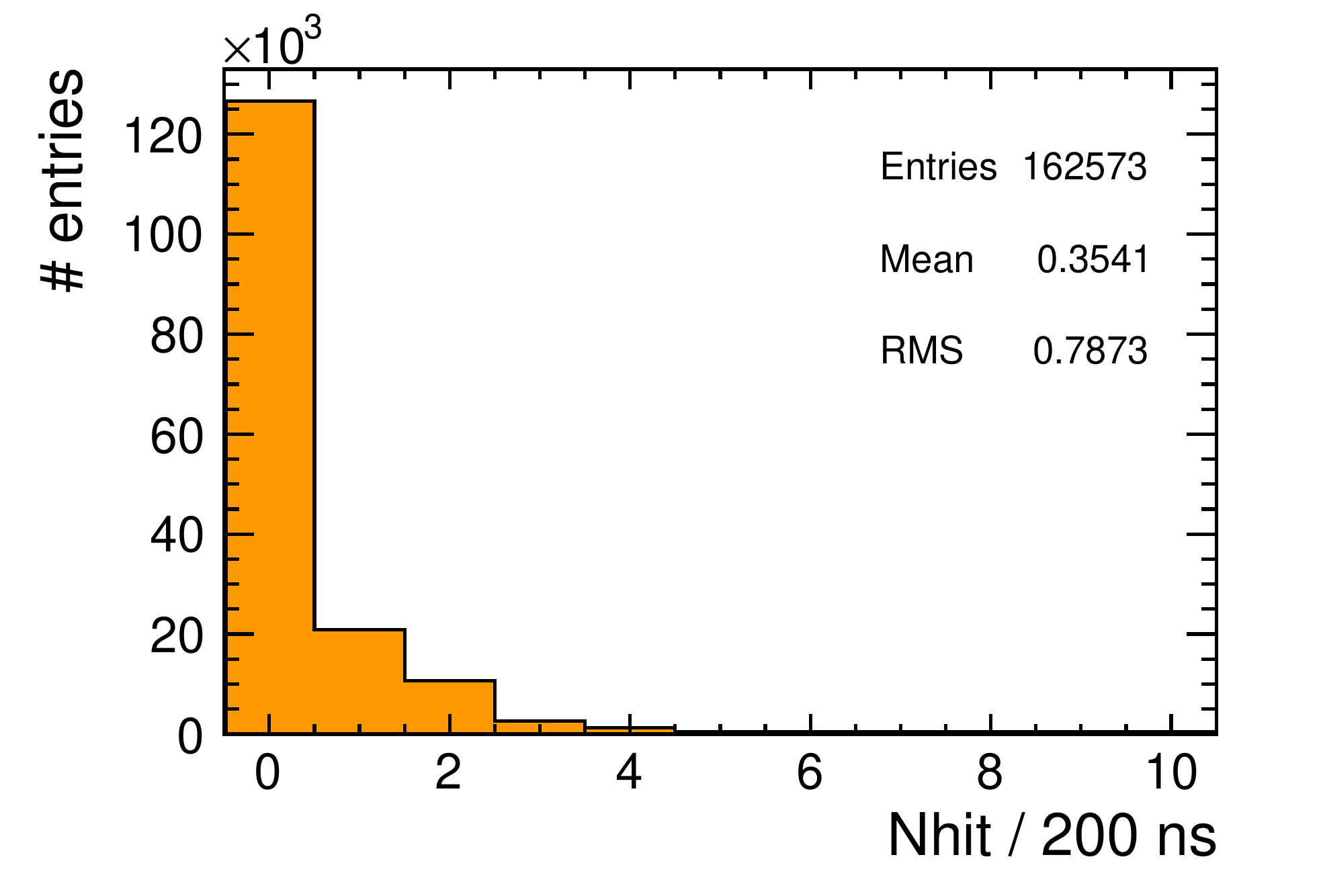}
    \caption{Distribution of the number of noise hits in a time slot of 200 ns for the whole detector.}
    \label{figs.noise_distribution}
  \end{center}
\end{figure}

\begin{figure}[!ht]
  \begin{center}
    \includegraphics[width=.65\textwidth]{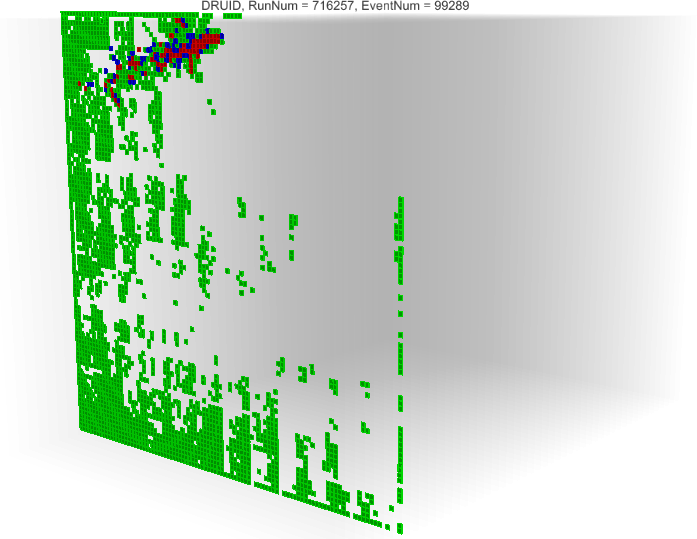}
    \caption{Example of coherent noise event display due to grounding problems. The three thresholds are respectively represented in green, blue and red.}
    \label{figs.grounding_display}
  \end{center}
\end{figure}

Because of the use of the trigger-less mode, an event building procedure is necessary. Indeed, with this mode, data collected contain signal from the beam particles as well as the cosmic rays and electronic noise due to gain fluctuation in the chambers. The event building procedure is based on a time clustering method. The time of each hit is recorded through the internal time counter of the ASIC. This clock is a 200 ns time slot. To build physical events (cosmics and beam particles) a histogram of the hit time is built (see Fig.\ \ref{figs.time_spectrum}). A time clusterisation is done when one time bin is found with more than 7 hits. Then if the selected hit time bin is a local maximum, hits in the two adjacent time bins are gathered inside.
Rejected bins are used to estimate the number of noise hits per time slot. From the noise distribution, shown in Fig.\ \ref{figs.noise_distribution}, the number of noise hits in physical events should not be larger than one. 
Among those selected physical events, it can still be found some events which are clearly due to electronic noise related to grounding problems. These events are easy to distinguish because of their particular topology. Indeed, for these events, number of hits in one layer is really larger than in others as it is shown in Fig.\ \ref{figs.grounding_display}.

\subsection{Detector performance}
\label{sec:building-perf-perf}
\begin{figure}[!ht]
  \begin{center}
    \includegraphics[width=.4\textwidth]{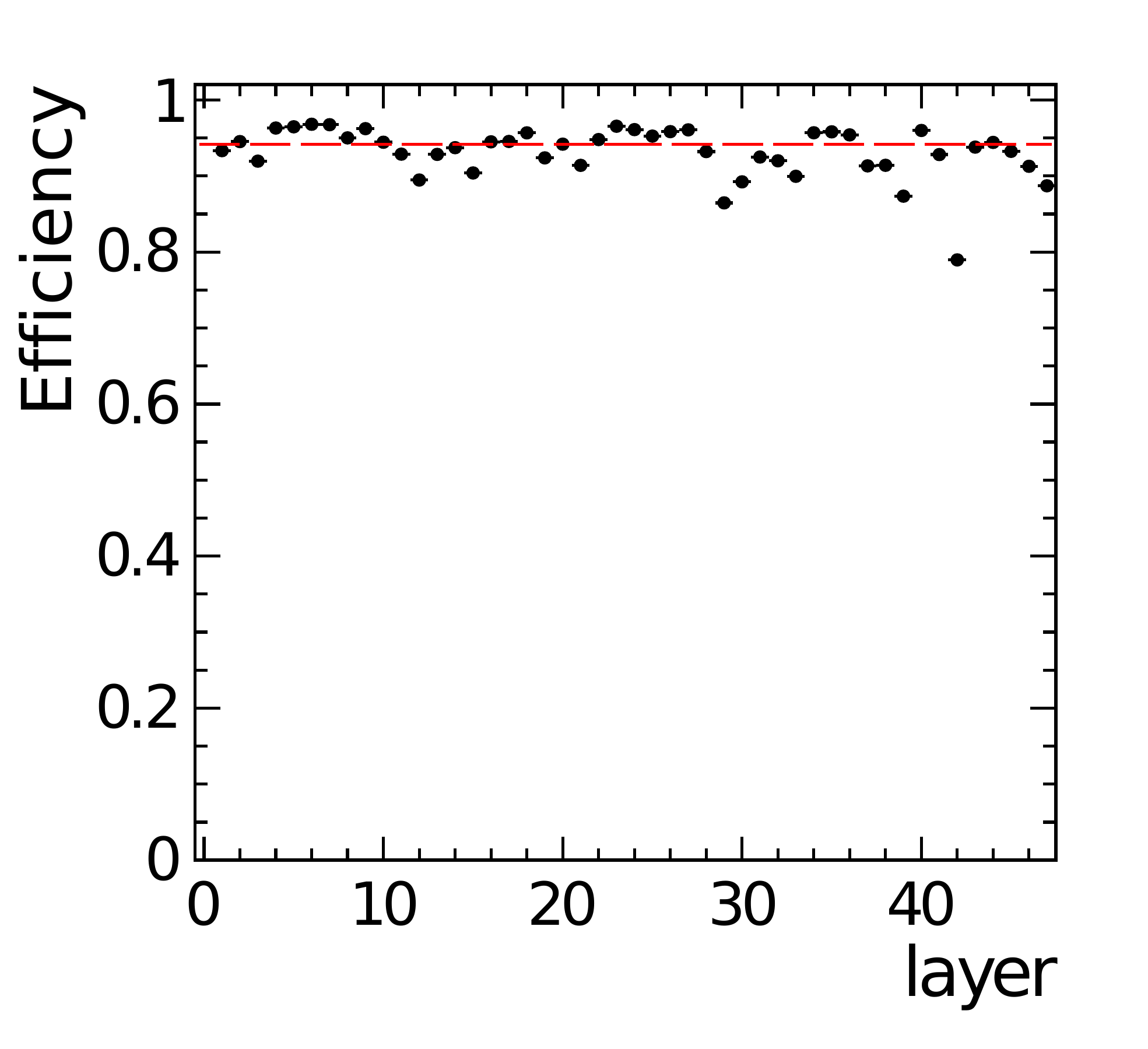}
    \includegraphics[width=.4\textwidth]{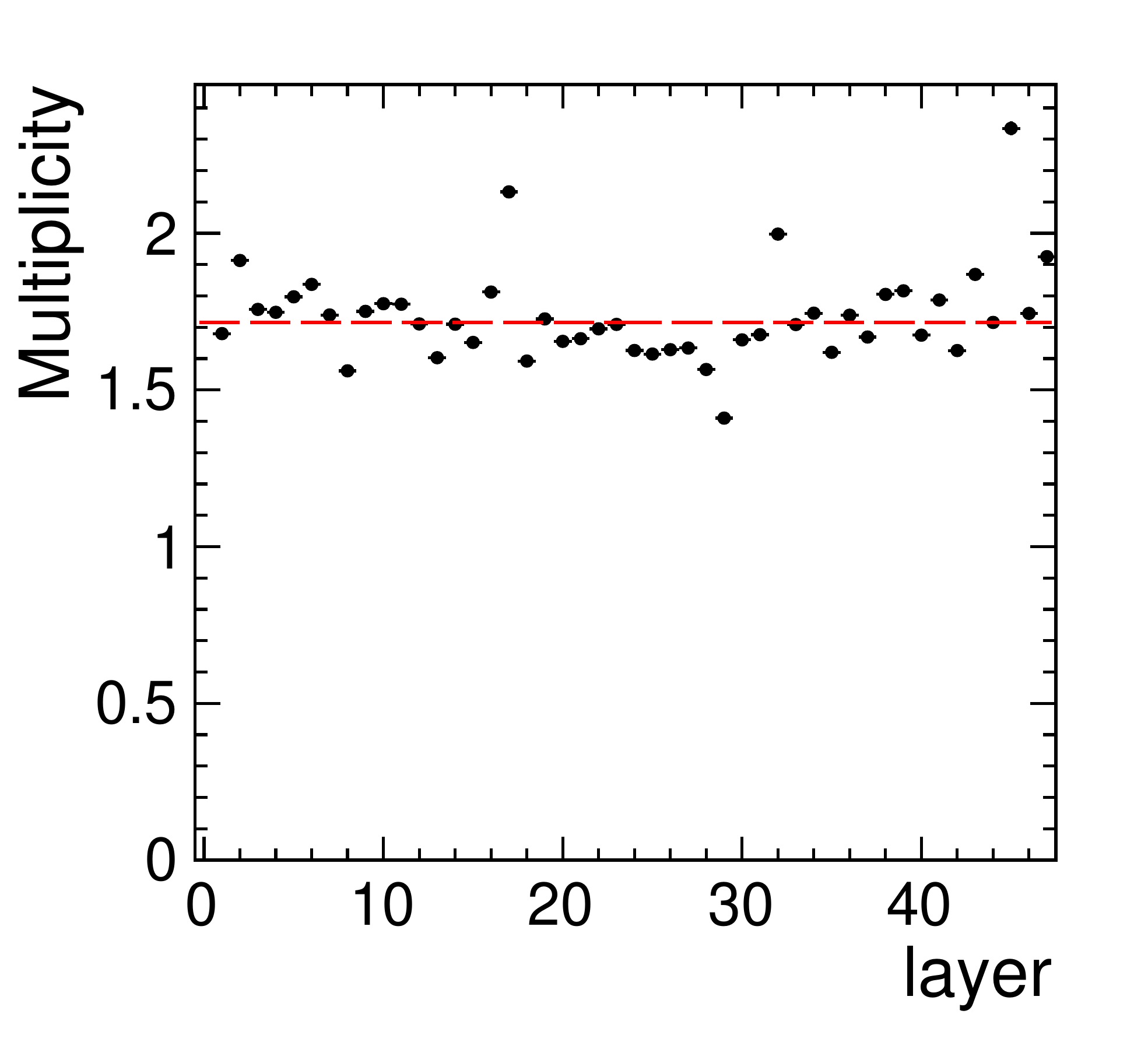}
    \caption{Effiency per layer (left). The red line represents the average efficiency ($\simeq 95 \%$). Multiplicity per layer (right). The red line represents the average multiplicity ($\simeq 1.73$).}
    \label{figs.efficiency_multiplicity}
  \end{center}
\end{figure}

Performance of the detector is estimated both online and offline. During the data acquisition a monitoring program was running to estimate the detector efficiency, the ASICs occupancy etc. This is very helpful to find out some detector problems and to check if the beam settings are optimal for the SDHCAL prototype. The efficiency and the particle multiplicity per layer are also estimated offline using the beam muons and cosmics. To study the efficiency of one layer, tracks reconstructed from other layers are used. To build such a track, hits in each layer are grouped into clusters using a nearest-neighbour approach. Isolated clusters are removed if the transverse distance to other clusters in the same layer and in other layers is higher than 12 cm. Track is then constructed if not more than one cluster per fired layer remains. The $\chi^2$ of the reconstructed track is estimated and the track is kept only if its $\chi^2$ is lower than 20. The expected track impact of the track in the studied layer is determined. This layer is counted as efficient if one cluster is found inside, at a distance lower than 2.5 cm with respect to the expected impact. The multiplicity is defined as the number of hits in the cluster if any. The efficiency and multiplicity per layer are shown in Fig.\ \ref{figs.efficiency_multiplicity}.


\section{Energy resolution of the SDHCAL prototype}
\label{sec:result}

\subsection{Particle identification}
\label{sec:result-identification}
The H6 beam line is contaminated by different kind of particles. Data samples contain also signal from cosmic particles because of the use of the trigger-less mode. Moreover no Cherenkov counter was used to discriminate pions against other particles. Those reasons make it necessary to find topological variables which will allow a powerful pion shower selection.

\subsubsection{Neutral and multi-particles contamination}
The rejection of neutral particles is done by requesting that the number of hits in the first five layers is exceeding 5. This will also reject a large part of cosmics. To eliminate multi-particles in the data samples the events in which the hit dispersion in each of the first five layers is greater than 5 cm are dropped. 
\label{sec:result-identification-neutral}

\subsubsection{Proton contamination}
\label{sec:result-identification-proton}
ATLAS Collaboration measured a quite high proton contamination in the H6 pion beam line above 20 GeV \cite{atlas-paper}.The proton interaction length is slightly lower than the one for pion which means that the number of hits for hadronic shower induced by protons have slightly more hits than the one from pions in the SDHCAL prototype. Consequently longitudinal leakage would be less important for protons and this effect is increasing with the energy. However, in the following protons are treated as pions.

\subsubsection{Muon contamination}
\label{sec:result-identification-muon}
Muons are present in the beam. They are coming from the pion interactions with the collimators along the beam line, from the decaying pions in front of the detector and from cosmics. To eliminate these muons it is requested that the ratio between the number of hits and the number of fired layers is larger than 2.2. This value is higher than the pad mean multiplicity which was found to be 1.73 (see Section \ref{sec:building-perf-perf}). In addition radiative muons are rejected by requesting that the ratio between the number of layers in which the hit dispersion is higher than 5 cm and the total number of fired layers is greater than 20$\%$. This will also rejects pions which start their shower in the ten last layers. In this case, the depth crossing by the pion before it starts its shower is about  $4.5\lambda_I$. The number of those pions should be negligible.

\subsubsection{Electron contamination}
\begin{figure}[!ht]
  \begin{center}
    \includegraphics[width=0.49\textwidth]{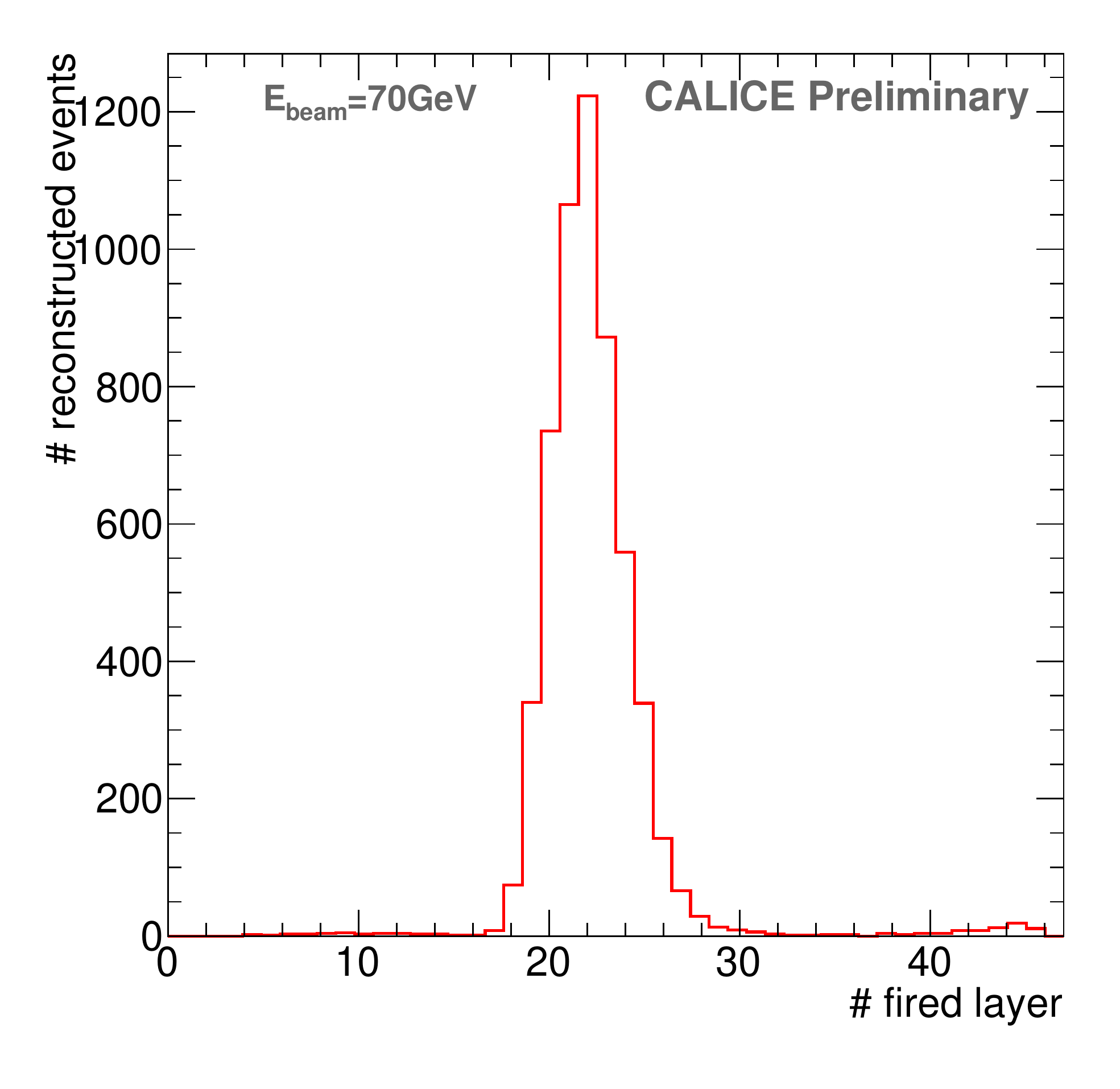}  
    \includegraphics[width=0.49\textwidth]{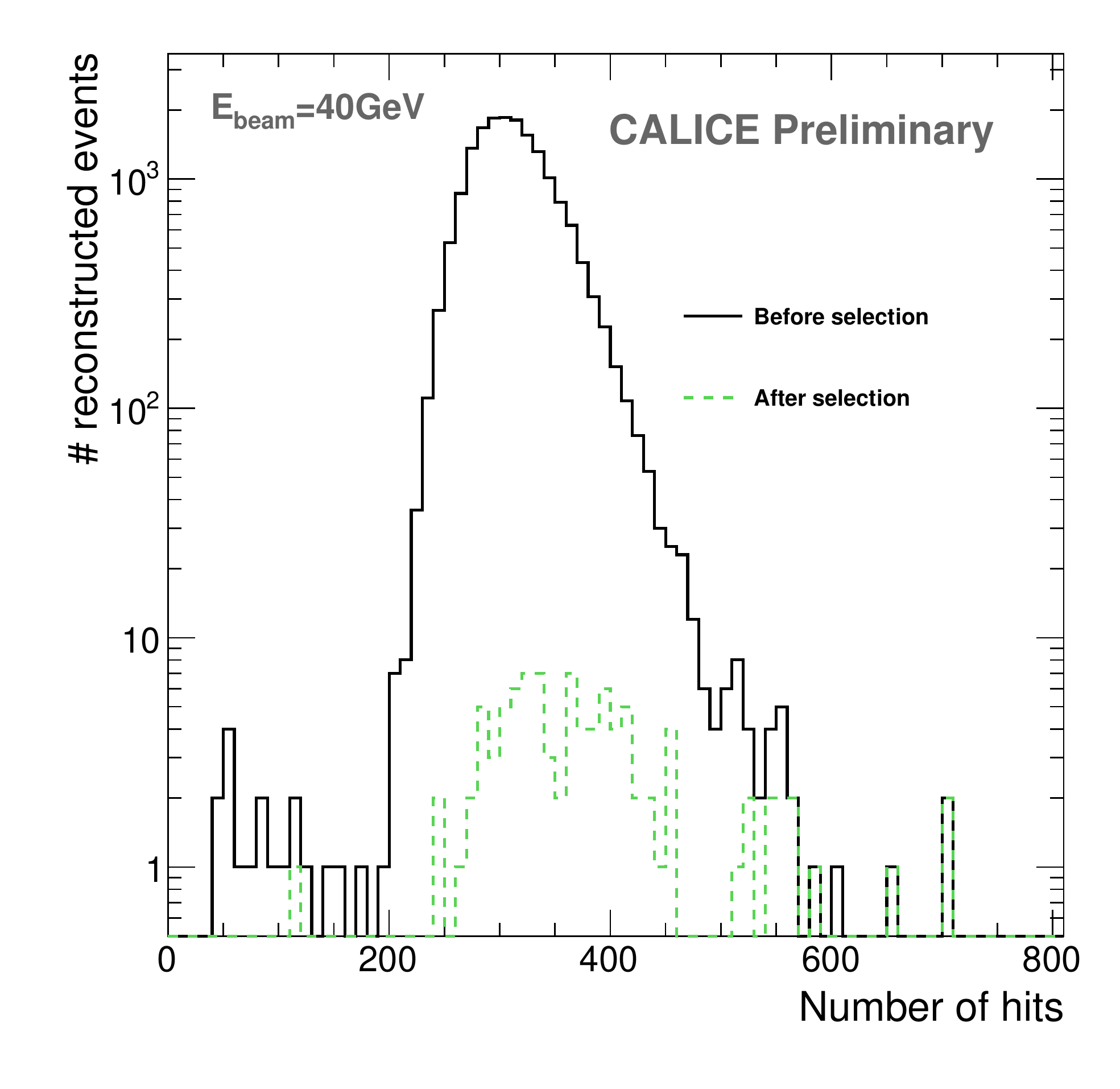}
    \caption{Distribution of the number of fired layers for a 70 GeV electron run after muon rejection (left). Distribution of number of hits for a 40 GeV electron run (right). The solid black line is the distribution before the electron rejection (after muon rejection) and the dotted green line is the one after.}
    \label{figs.electron-cut}
  \end{center}
\end{figure}

Electrons are also present in the beam even though the presence of a 4 mm thick lead absorber is supposed to eliminate most of them. The contamination is rather important at energies below 20 GeV. To eliminate electrons, the fact that these particles start their shower in the first plates is used. Indeed radiation length $X_0$ in the steel is about 1.75 cm. So in order to reject electromagnetic showers in the data samples, events in which the shower is starting before the fifth layer are rejected. The corresponding depth is about $6X_0$. The shower starting layer (SSL) is then defined as the first layer with at least 4 hits and with the three following layers having at least 4 hits. Electromagnetic showers in the energy range between 5 and 80 GeV are longitudinally contained in 30 layers as it is shown in Fig.\ \ref{figs.electron-cut}. To avoid the rejection of many hadron showers, the previous criterion (SSL$\geq$5) is only applied if the total number of fired layers is less than 30. With this, the number of rejected hadron showers at high energy will be negligible. At low energy, the part of rejected hadron showers is more important, but the effect of this selection is only statistical, and the energy resolution is not affected. The rejection power of this selection was checked on electron runs, and shown in Fig.\ \ref{figs.electron-cut}.
\label{sec:result-identification-electron}
\subsubsection{Pion selection summary}
\begin{figure}[!ht]
  \begin{center}
    \includegraphics[width=0.49\textwidth]{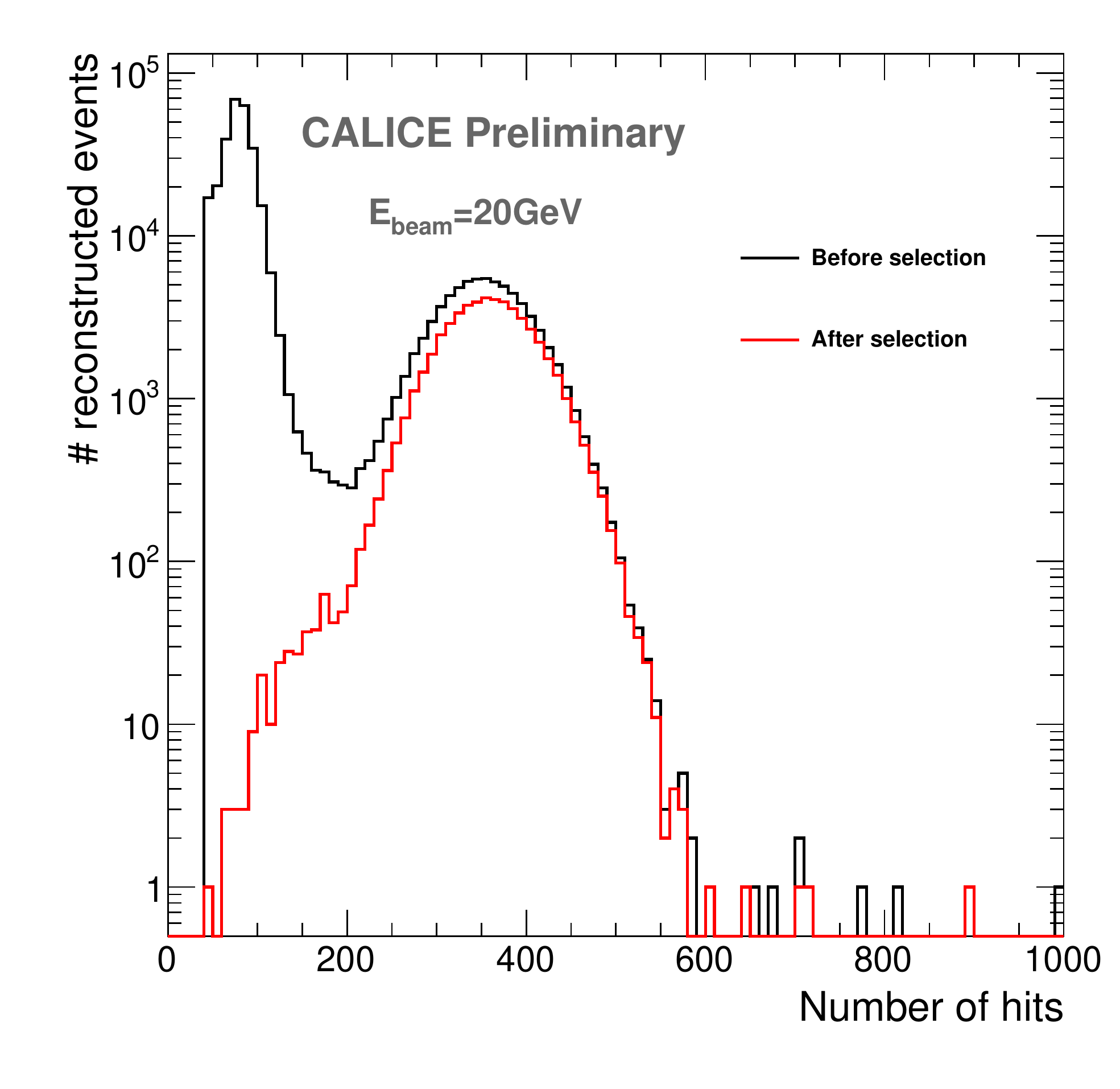}  
    \includegraphics[width=0.49\textwidth]{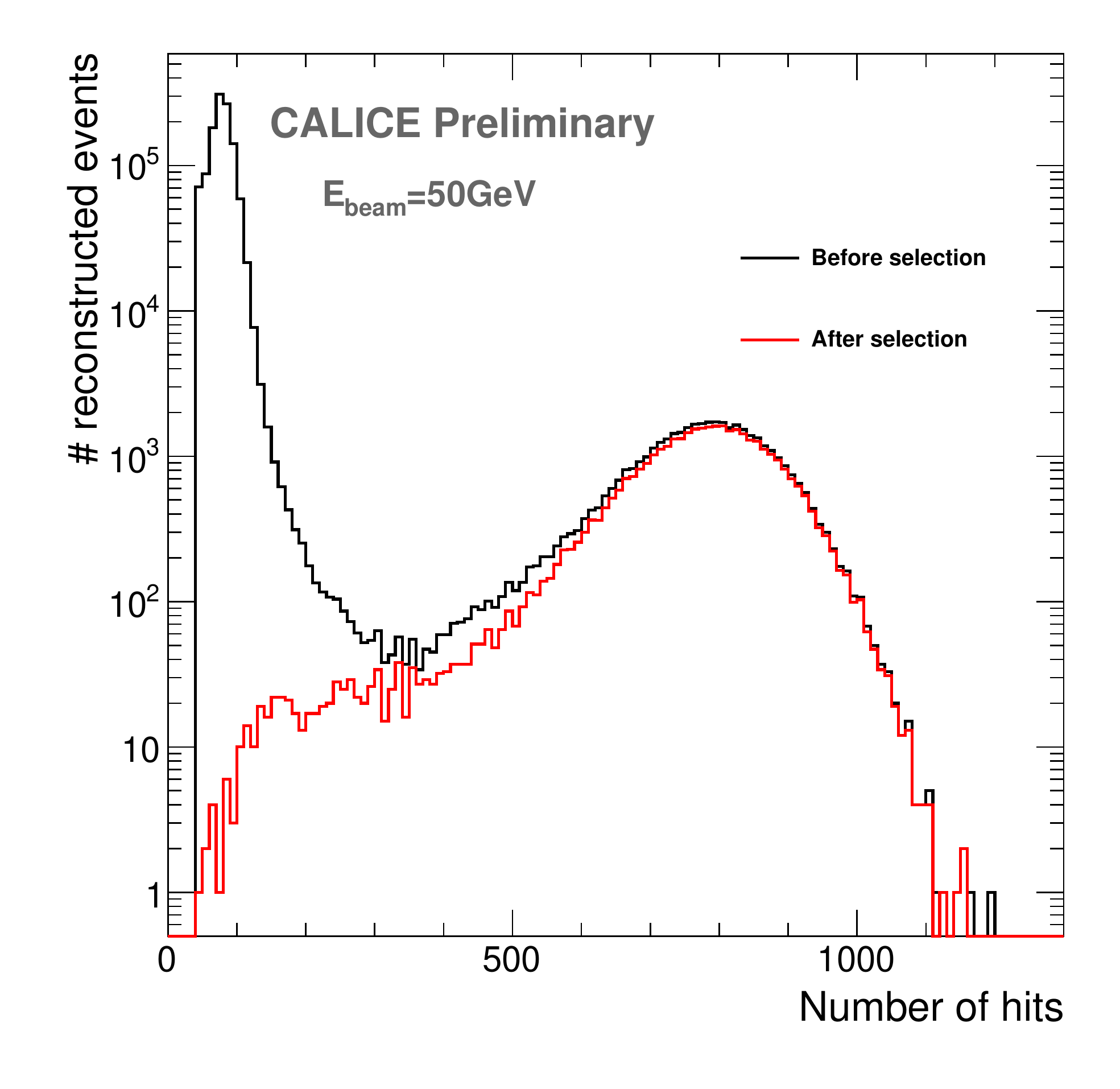}
    \caption{Number of hit distribution for 20 (left) and 50 (right) GeV pion runs before (black line) and after (red line) the whole selection}
    \label{figs.pion-selection}
  \end{center}
\end{figure}

The cuts applied on data sample to select hadron showers are summarized in Table~\ref{tab.cut-summary}. Figure~\ref{figs.pion-selection} shows the results of this selection for 20 and 50 GeV pion runs.
\begin{table}[htp]
  \begin{center}
    \begin{tabular}{c|c}
      Description & Variables and selection\\
      \hline
      Muon rejection & $\frac{N_{hit}}{N_{layer}}>2.2$\\
      Neutral rejection &$N_{hit\in  First\ 5\ layers} \geq4$\\
      Radiative muon rejection&$\frac{N_{layer}\backslash RMS>5cm}{N_{layer}}>20\%$\\
      Electron rejection &$SSL\ \geq 5\ OR\ N_{layer} \geq 30$\\
    \end{tabular}
  \end{center}
  \caption{Summary of different cuts applied on data samples to select pions}
  \label{tab.cut-summary}
\end{table}


\subsection{Energy resolution}
\label{sec:result-energy}
\begin{figure}[!ht]
  \begin{center}
    \includegraphics[width=.7\textwidth]{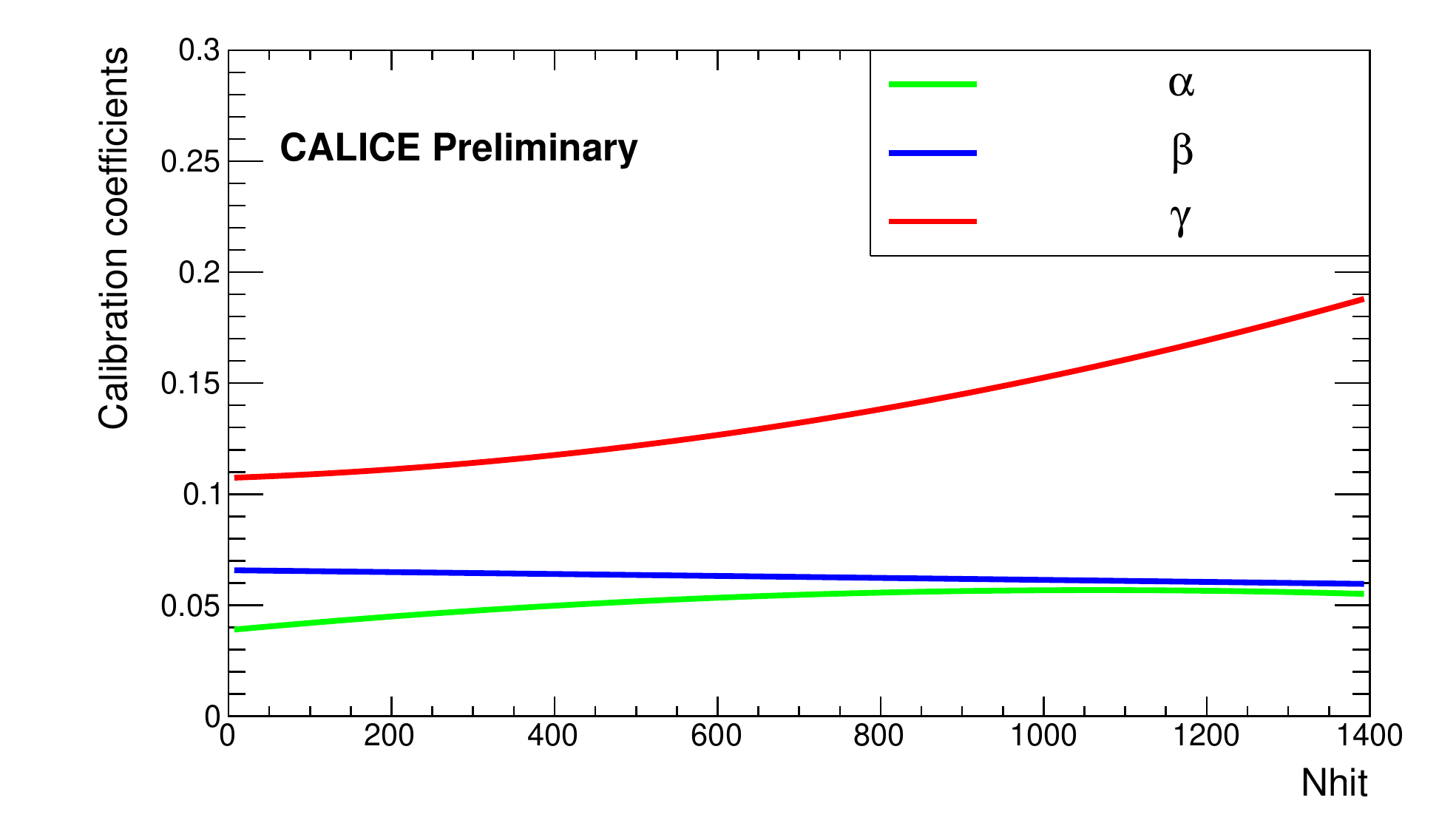}
    \caption{$\alpha$ (green line), $\beta$ (blue line) and $\gamma$ (red line) evolution in terms of the total number of hits.}
    \label{figs.params-evolution}
  \end{center}
\end{figure}

\begin{figure}
  \includegraphics[width=.33\textwidth]{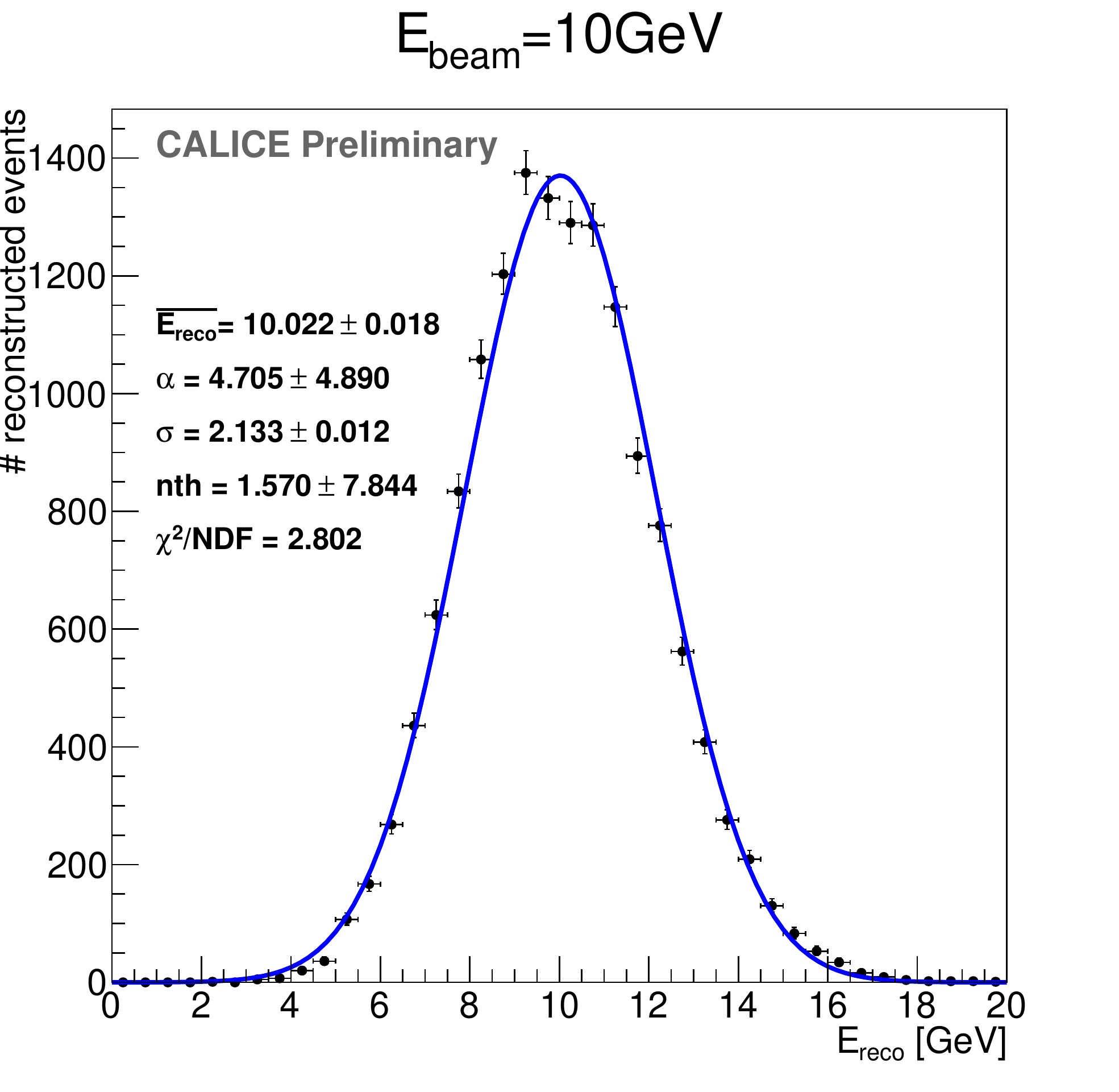}
  \includegraphics[width=.33\textwidth]{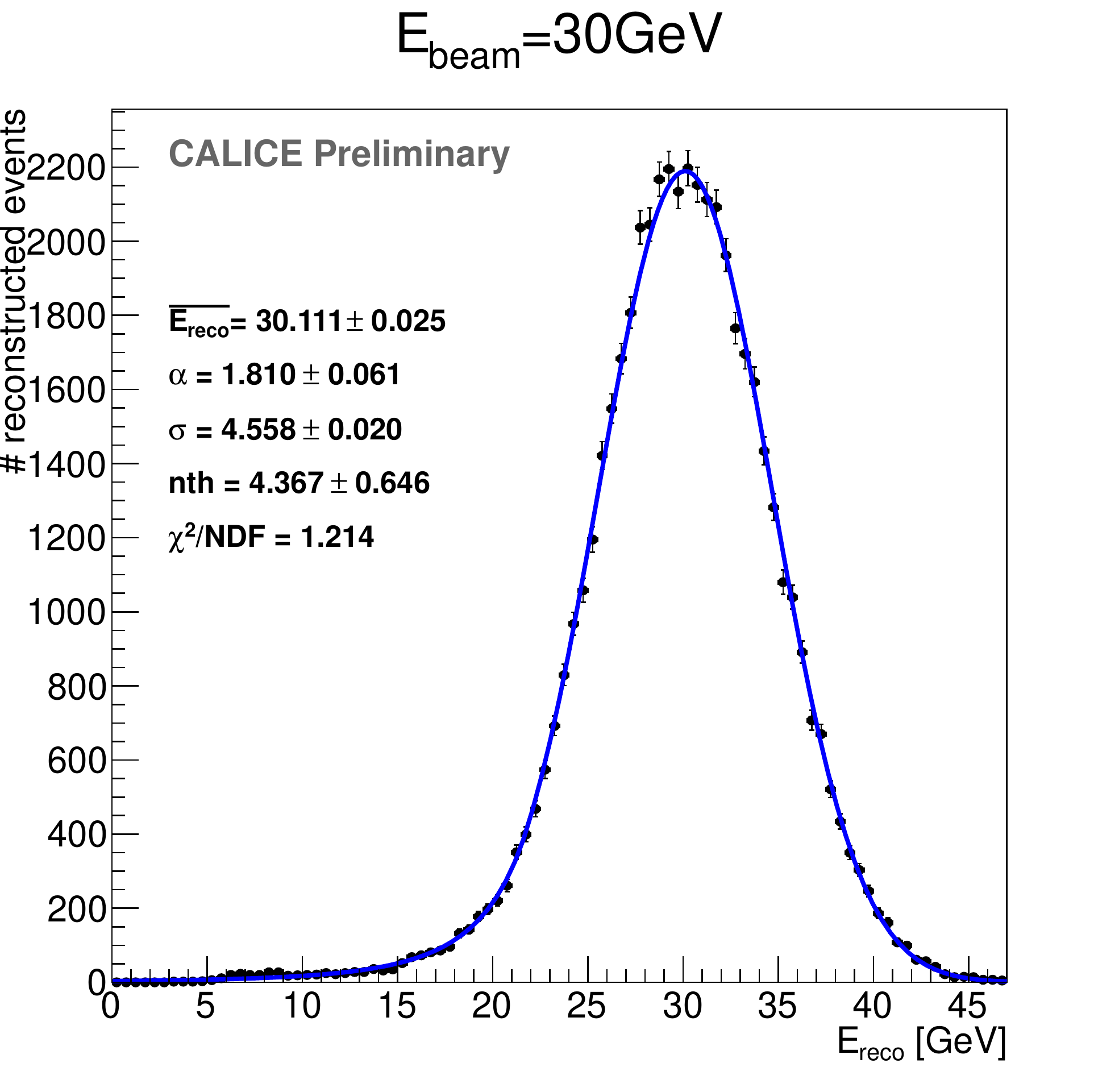}
  \includegraphics[width=.33\textwidth]{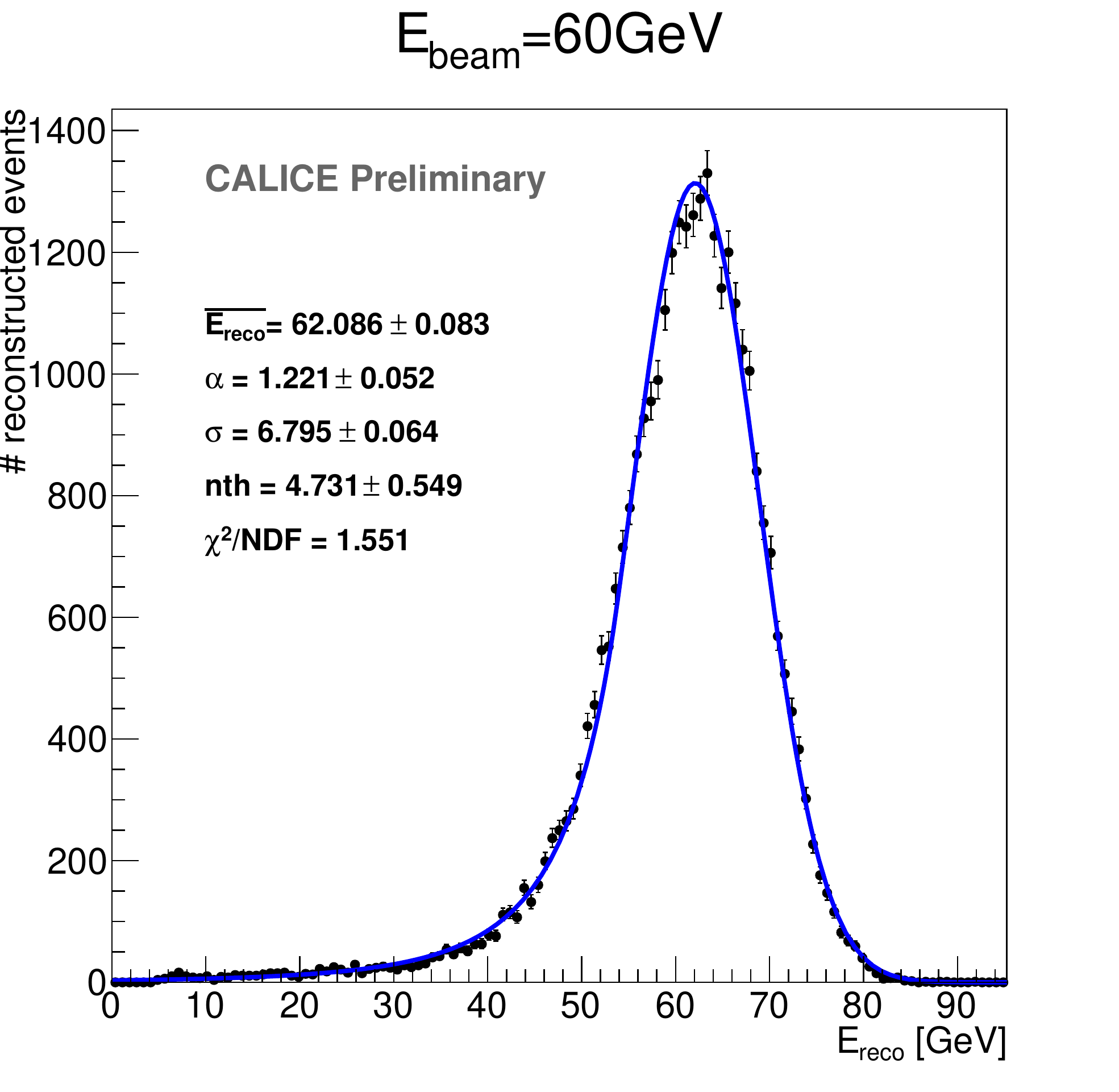}
  \caption{Reconstructed energy distributions for 10, 30 and 60 GeV pion runs.}
  \label{figs.energy-distribution}
\end{figure}

\begin{figure}
  \includegraphics[width=.5\textwidth]{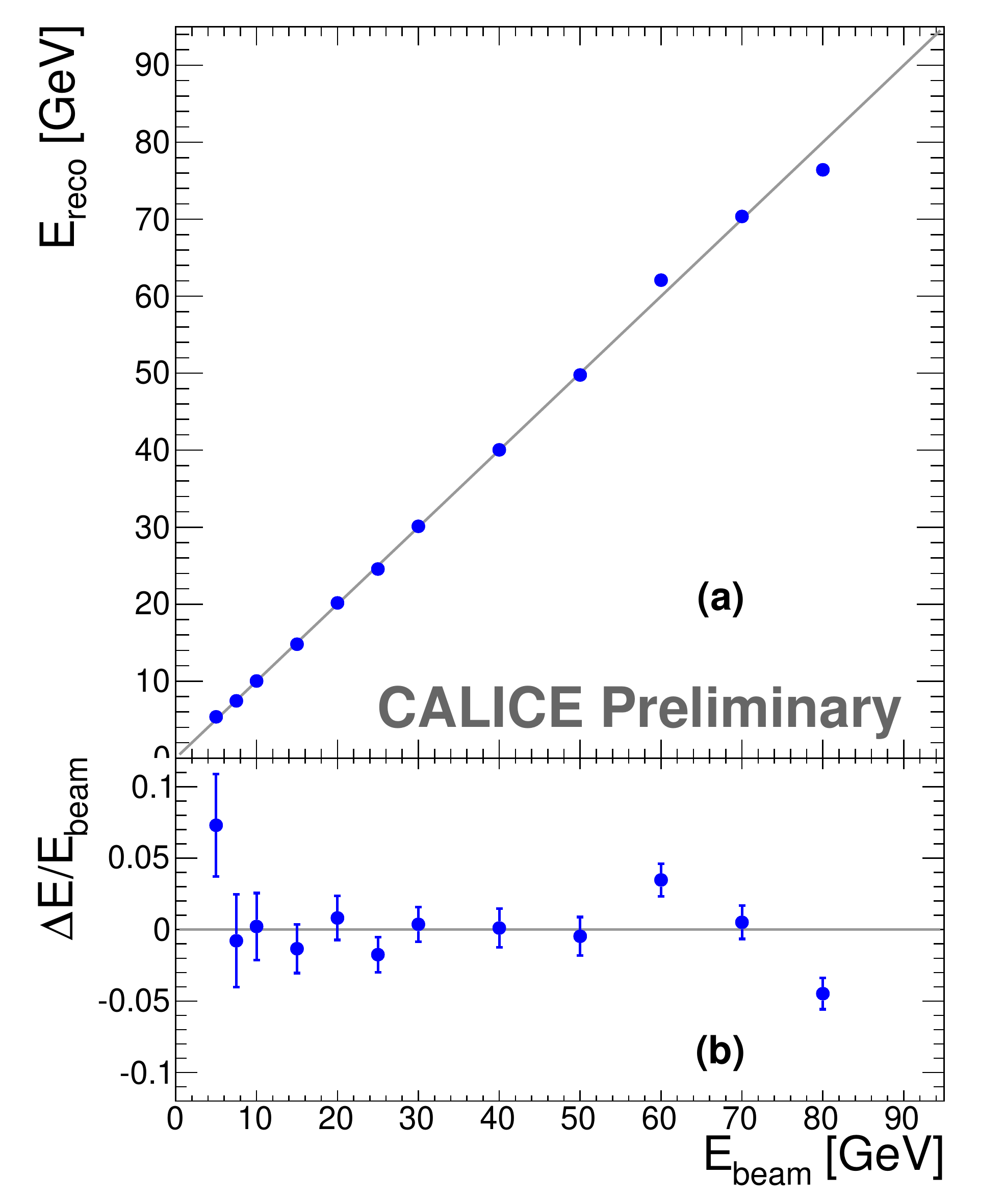}
  \includegraphics[width=.5\textwidth]{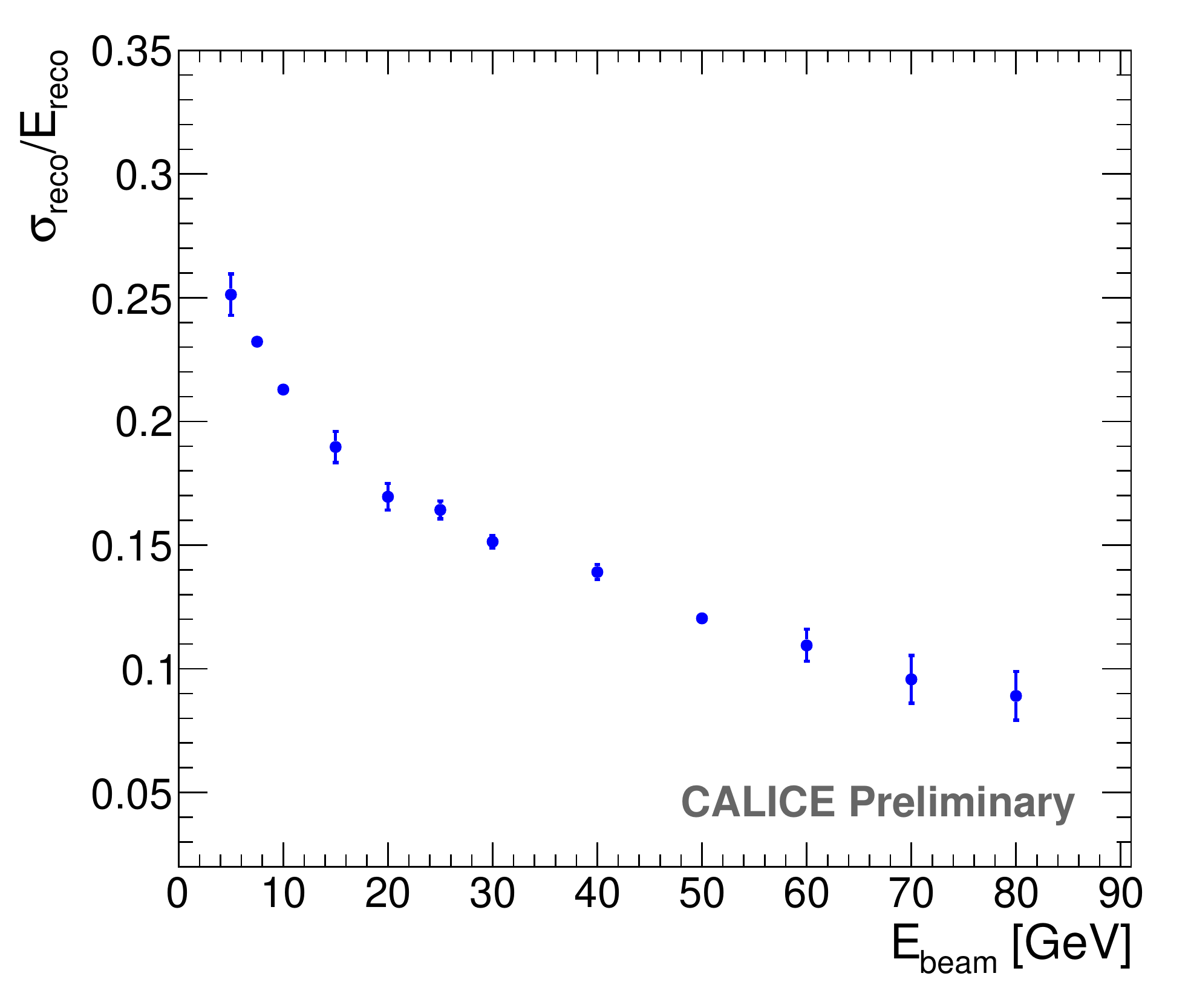}
  \caption{left: (a): Mean reconstructed energy for pion showers and (b): relative deviation of the pion mean reconstructed energy with respect to the beam energy as a function of the beam energy. right: $\frac{\sigma(E)}{E}$ of the reconstructed pion energy E as a function of the beam energy (right). The reconstructed energy is computed using the three thresholds information (see Eq. \protect\ref{eq.energy-reco}) and the distributions are fitted with a Crystal Ball function.}
  \label{figs.results}
\end{figure}

The reconstructed energy is calculated after applying the previous selection with the same method described in \cite{sdhcal-paper}. The reconstructed energy is given by equation \ref{eq.energy-reco}.\\
\begin{equation}   
  E_{reco}=\alpha(N_{hit}) N_1 + \beta(N_{hit}) N_2 + \gamma(N_{hit}) N_3
  \label{eq.energy-reco}
\end{equation}
The coefficients $\alpha$, $\beta$, $\gamma$ are quadratic functions of total number of hits and $N_i$ is the number of hits for the $i^{th}$ threshold. The choice of such a parametrization comes from a previous study which indicates that the quadratic parametrization is needed to restore the linearity \cite{sdhcal-paper}. These coefficients are extracted using a $\chi^2$ minimisation (see equation \ref{eq.chi2-minim}).\\ 
\begin{equation}
  \chi^2 = \sum_{i=1}^N{\frac{(E_{beam}-E_{reco})^2}{E_{beam}}}
  \label{eq.chi2-minim}
\end{equation}
N is the number of events used for the minimisation. The evolution of these parameters is shown in Fig. \ref{figs.params-evolution}.
These coefficients are then used to estimate the energy of incoming particles. Fig. \ref{figs.energy-distribution} shows the energy distributions for 10, 30 and 60 GeV pions. The distributions are fitted with the Crystal Ball function defined in \cite{sdhcal-paper} in order to take into consideration the tail at low energy. A gaussian fit was also performed and used for estimation of systematic uncertainties.
The mean reconstructed energy for pion showers and the relative deviation of the pion mean reconstructed energy with respect to the beam energy are shown in Fig. \ref{figs.results}. The relative deviation of the pion mean reconstructed energy with respect to the beam energy is lower than 5$\%$ on a large energy scale ([7.5-80] GeV). Figure \ref{figs.results} shows the hadronic energy resolution of the SDHCAL prototype which reaches 9$\%$ at 80 GeV. Since no correction was applied these results are very encouraging. 


\section{Conclusion}
The SDHCAL prototype achieved in 2011 was successfully tested at the SPS in CERN in 2012 using trigger-less mode and power pulsed mode. It has been exposed to beams of pions, electrons and muons on a large energy range (5-80 GeV). Encouraging results on the energy reconstruction and resolution have been reached without any gain correction. New algorithms to linearize the pion response are developed. They allow to obtain a satisfactory linearity on the energy reconstruction and a good energy resolution which reaches 9$\%$ at 80 GeV. 
\label{conclusion}



\end{document}